\documentclass{article}
\usepackage{spconf,amsmath,graphicx,hyperref}
\usepackage[table]{xcolor}
\usepackage{comment}
\usepackage{multirow}
\usepackage{arydshln}
\usepackage{subfigure}
\usepackage{algorithm}
\usepackage{algorithmic}
\usepackage{cite}
\usepackage{makecell}
\usepackage{booktabs}
\usepackage{mwe}
\usepackage{enumitem}

\definecolor{deepred}{rgb}{0.698,0.133,0.133}
\definecolor{deepgreen}{rgb}{0.23,0.49,0.14}
\definecolor{LightCyan1}{rgb}{0.88,1,1}
\definecolor{shadecolor}{rgb}{0.92,0.92,0.92}

\makeatletter
\renewcommand{\@fnsymbol}[1]{\ensuremath{\ifcase#1\or \dagger\or \ddagger\or \S\or \P\or \#\or \|\or **\or \dagger\dagger\or \ddagger\ddagger\else *\fi}}
\makeatother

\title{OV-InstructTTS: Towards Open-Vocabulary Instruct Text-to-Speech}
\name{
  \begin{tabular}{c}
    Yong Ren$^{12}$ 
    Jiangyan Yi$^{3}$\sthanks{Corresponding author}, 
    Jianhua Tao$^{34\dagger}$, 
    Haiyang Sun$^{1}$, 
    Zhengqi Wen$^{4\dagger}$, 
    Hao Gu$^{12}$,
    Le Xu$^{1}$, 
    Ye Bai$^{1}$
  \end{tabular}
}

\address{
\textsuperscript{1}The State Key Laboratory of Multimodal Artificial Intelligence Systems, \\Institute of Automation, Chinese Academy of Sciences, China \\
\textsuperscript{2}School of Artificial Intelligence, University of Chinese Academy of Sciences, China \\
\textsuperscript{3}Department of Automation, Tsinghua University, China \\
\textsuperscript{4}Beijing National Research Center for Information Science and Technology, Tsinghua University, China \\
}

\begin{document}
\ninept
\maketitle
\begin{abstract}
Instruct Text-to-Speech (InstructTTS) leverages natural language descriptions as style prompts to guide speech synthesis. However, existing InstructTTS methods mainly rely on a direct combination of audio-related labels or their diverse rephrasings, making it difficult to handle flexible, high-level instructions. Such rigid control is insufficient for users such as content creators who wish to steer generation with descriptive instructions. To address these constraints, we introduce \textbf{OV-InstructTTS}, a new paradigm for open-vocabulary InstructTTS. We propose a comprehensive solution comprising a newly curated dataset, OV-Speech, and a novel reasoning-driven framework. The OV-Speech dataset pairs speech with open-vocabulary instructions, each augmented with a reasoning process that connects high-level instructions to acoustic features. The reasoning-driven framework infers emotional, acoustic, and paralinguistic information from open-vocabulary instructions before synthesizing speech. Evaluations show that this reasoning-driven approach significantly improves instruction-following fidelity and speech expressiveness. We believe this work can inspire the next user-friendly InstructTTS systems with stronger generalization and real-world applicability. The dataset and demos are publicly available on our project page.\footnote{https://y-ren16.github.io/OV-InstructTTS} 
\end{abstract}
\section{Introduction}
\label{sec:intro}
In recent years, text-to-speech (TTS) technology has achieved remarkable progress. Advanced TTS systems have reached human-level speech quality and demonstrated the ability to clone arbitrary speaker timbres from a short audio sample\cite{wang2023neural, wang2025maskgct, liao2024fish, anastassiou2024seed, du2024cosyvoice, chen2024f5, zhou2025indextts2}. The rapid development and accessibility of these systems have accelerated content creation, personalization, and human–computer interaction. As TTS technology advances toward real-world deployment, enhancing TTS customizability has become a key research focus \cite{xie2024towards}.

To improve TTS controllability, recent research has explored using natural language descriptions as instructions, known as InstructTTS \cite{guo2023prompttts,yang2024instructtts,leng2024prompttts,ji2024textrolspeech,jin2024speechcraft,zhou2024voxinstruct,yang2025emovoice}. PromptTTS \cite{guo2023prompttts} first introduced style prompts based on five predefined attributes. This was later extended by InstructTTS \cite{yang2024instructtts}, which utilized longer and more descriptive sentences. More recent systems, such as PromptTTS2 \cite{leng2024prompttts}, TextrolSpeech \cite{ji2024textrolspeech}, and SpeechCraft\cite{jin2024speechcraft}, utilize large language models (LLMs) to generate detailed and diverse style prompts. 
Furthermore, VoxInstruct \cite{zhou2024voxinstruct} shifted toward better user alignment by replacing separate content and style prompts with free-form human instructions, capturing the user’s intent more holistically.

However, existing InstructTTS systems remain fundamentally constrained by their reliance on direct acoustic attributes (e.g., pitch, speaking rate, and emotion). Instructions in these systems are typically derived from the combination or diverse rephrasing of predefined acoustic labels. While effective for basic control, these approaches restrict creativity, as they fail to capture the complex mapping between high-level instructions and low-level acoustic features. As noted in SpeakEasy \cite{brade2025speakeasy}, content creators still lack intuitive ways to convey their expressive intent to TTS models. Most recently, InstructTTSEval \cite {huang2025instructttseval} introduced a Role-Play Instruct task, but it was limited to a small evaluation set and did not provide a solution.

In this paper, we introduce a novel paradigm \textbf{OV-InstructTTS}, which aims to make InstructTTS more user-friendly and flexible by enabling models to synthesize speech that aligns with users’ intentions directly from open-vocabulary instructions. To support this, we construct \textbf{OV-Speech}, a multi-layered dataset built upon the ContextSpeech corpus\footnote{https://huggingface.co/datasets/Insects/ContextSpeech}. A key feature of OV-Speech is that its instructions are derived from narrative context rather than reformulations of acoustic or emotional labels. This design ensures that the instructions are diverse and unconstrained in expression, while remaining consistent with the target speech. To bridge the gap between high-level instructions and low-level acoustics, we augment each sample with a reasoning chain generated by LLMs. Furthermore, we enrich transcriptions with paralinguistic tags. 

Building on the OV-Speech dataset, we further propose a reasoning-driven framework OV-InstructTTS-Thinking-Emotion-Paralanguage (\textbf{OV-InstructTTS-TEP}) for the OV-InstructTTS task. The primary challenge lies in bridging the significant semantic gap between high-level, open-vocabulary instructions and their low-level acoustic realizations. Recent advancements in large audio language models(LALMs), such as Kimi-Audio \cite{ding2025kimi}, Qwen2.5-Omini \cite{xu2025qwen2}, and Step-Audio 2 \cite{wu2025step}, have demonstrated powerful unified capabilities in comprehension, reasoning, and generation. These characteristics make LALMs particularly well-suited for OV-InstructTTS, as their reasoning abilities enable them to deduce appropriate vocal styles from open-ended instructions before synthesis. We build a reasoning-driven system that reframes OV-InstructTTS into a new paradigm: it first derives explicit emotional and acoustic descriptions from the open-vocabulary instruction via a thinking process. Subsequently, it generates an interleaved sequence of audio and text tokens—where the text is enriched with emotional labels and paralinguistic tags. 
Our key contributions are as follows:
\begin{itemize}[leftmargin=*]
    \item \textbf{Paradigm:} This paper proposes OV-InstructTTS, a novel paradigm that shifts instructTTS beyond its dependency on rephrased audio attributes, pushing controllable speech synthesis towards more flexible and user-friendly real-world applications.
    \item \textbf{Dataset:} We construct OV-Speech, a large-scale dataset providing a foundation for this paradigm. It features open-vocabulary instructions derived from narrative context, reasoning chains that connect instructions to acoustics, and transcriptions enriched with paralinguistic tags.
    \item \textbf{Method:} This paper proposes OV-InstructTTS-TEP, a novel reasoning-driven OV-InstructTTS framework based on LALM. Our method is designed to interpret open-ended instructions through a reasoning process to generate highly expressive speech that is consistent with the user's intent.
    \item \textbf{Experiments:} Extensive experiments and ablation studies demonstrate the value of our dataset and the effectiveness of the OV-InstructTTS-TEP framework. LLM-a-a-judge and subjective evaluations confirm the consistency of synthesized speech with open-ended instructions.
\end{itemize}

\section{The OV-Speech Dataset Construction}
\label{sec:dataset}
\begin{figure}[t]
  \centering
  \includegraphics[width=\linewidth]{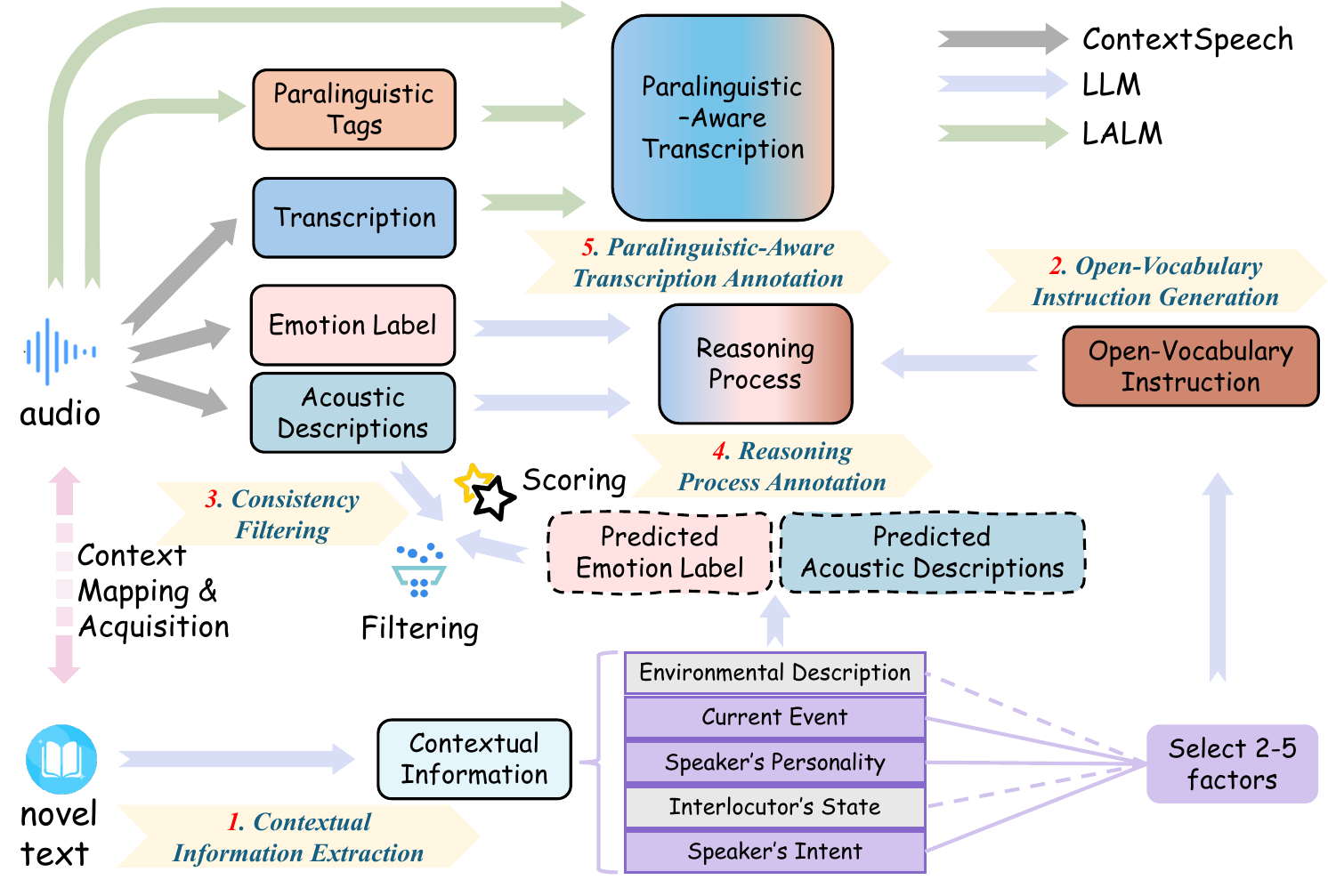}
  \vspace{-5pt}
  \caption{The pipeline for constructing the OV-Speech dataset.}
  \label{fig:dataset}
\end{figure}

The first challenge in developing an OV-InstructTTS system is the lack of suitable data that pairs speech with open-vocabulary instructions. To address this, we constructed OV-Speech, a large-scale, multi-layered dataset built upon the ContextSpeech corpus, a 476.8-hour resource of multi-speaker audiobooks paired with their corresponding novels. From this foundation, we leverage its single-sentence audio utterances, text transcriptions, and existing emotional labels and acoustic descriptions. As illustrated in Fig. \ref{fig:dataset}, the construction of OV-Speech follows a five-stage pipeline.

\textbf{Contextual Information Extraction.}
The foundation for our open-vocabulary instructions is the rich narrative surrounding each utterance. For each audio sample, we first align it with the source novel and extract a context window of 1000 words preceding and following the target utterance. We then use Qwen3-32B\footnote{https://huggingface.co/Qwen/Qwen3-32B} \cite{yang2025qwen3} with strong understanding and reasoning abilities to distill this raw text into a set of structured contextual elements, including environmental descriptions, the current event, the speaker's personality, the interlocutor's state, and the speaker's intent. This process effectively captures the nuanced situational information relevant to the speech performance.

\textbf{Open-Vocabulary Instruction Generation.}
We then generate diverse high-level instructions based on the extracted contextual elements. For each sample, 2–5 elements are randomly selected and provided to Qwen3-32B, which is prompted to synthesize an open-vocabulary instruction akin to what a director might give to a voice actor. This two-step process, involving both random selection and creative generation, ensures a wide variety of unconstrained and descriptive instructions that are grounded in the narrative.

\textbf{Consistency Filtering.}
A critical challenge is ensuring that the novel-narration-derived open-vocabulary instruction genuinely aligns with the acoustic style of the spoken audio. To this end, we adopt an LLM-as-a-Judge filtering strategy: Deepseek-R1 \cite{guo2025deepseek} is given the speaker identity, transcript, and contextual information, and tasked with predicting emotional and acoustic attributes. These predictions are compared with the ground-truth labels from ContextSpeech by an LLM judge Qwen3-32B, and samples with low alignment scores (emotion $<$ 6 or acoustics $<$ 5) are discarded.

\textbf{Reasoning Process Annotation.}
This core annotation step enables our reasoning-driven InstructTTS model. For each sample that passes the consistency filter, we task a Qwen3-32B with generating a step-by-step "reasoning chain" that connects the high-level instruction to the low-level acoustic attributes. This reasoning chain consists of two stages: first, Instruction Deconstruction, where the model identifies the key contextual elements implied by the instruction, and second, Attribute Inference, where it deduces the emotional label and acoustic descriptions from these elements.

\begin{table}[t]
\vspace{-15pt}
\caption{Evaluation of transcription with paralinguistic tags generation methods. \textcolor{deepred}{\textbf{Red}} highlights the best performance.} 
\label{tab:r0}
\centering
\resizebox{\columnwidth}{!}{%
\begin{tabular}{lccccc}
\toprule
& \multicolumn{4}{c}{\textbf{NVSpeech170k-Testset}} & \multicolumn{1}{c}{\textbf{OV-Speech}}\\
\cmidrule(lr){2-5} \cmidrule(lr){6-6}
\textbf{Method} & \textbf{C-F1}$\uparrow$ & \textbf{P-F1}$\uparrow$ & \textbf{APS}$\downarrow$ & \textbf{S-CER(\%)} $\downarrow$ & \textbf{S-CER(\%)} $\downarrow$ \\
\midrule
PASR & 0.87 & 0.60 & 7.72 & 18.91 & 12.95 \\
PRI  & \textcolor{deepred}{\textbf{0.88}} & \textcolor{deepred}{\textbf{0.83}} & \textcolor{deepred}{\textbf{6.56}} & 0.12 & 4.04 \\
PC-PTI & \textcolor{deepred}{\textbf{0.88}} & 0.82 & 6.67 & \textcolor{deepred}{\textbf{0.03}} & \textcolor{deepred}{\textbf{0.35}}\\
\bottomrule
\end{tabular}%
}
\vspace{-12pt}
\end{table}

\textbf{Paralinguistic-Aware Transcription Annotation.}
To provide the TTS model with a more informative prediction target, we enrich the raw text transcriptions with 18 paralinguistic tags from NVSpeech \cite{liao2025nvspeech} (e.g., \texttt{[Laughter]}, \texttt{[Cough]}). To achieve this, we fine-tuned the Qwen2-Audio-7B \cite{chu2024qwen2}\footnote{https://huggingface.co/Qwen/Qwen2-Audio-7B} model on the NVSpeech170k \footnote{https://huggingface.co/datasets/Hannie0813/NVSpeech170k} dataset. We explored three strategies for this task: Paralinguistic-Aware Speech Recognition (\textbf{PASR}): predicting transcription with paralinguistic tags directly from audio. Paralinguistic Recognition and Insert (\textbf{PRI}): taking both audio and raw transcription to predict the final tagged transcript. Paralinguistic Classification then Tag Insert (\textbf{PC-PTI}): A two-stage approach that first predicts the presence of paralinguistic events from audio, then inserts the corresponding tags into the correct positions within the raw transcript. As shown in Table~\ref{tab:r0}, we evaluate performance using \emph{Category F1-Score (C-F1)} for tag classification, \emph{Position F1-Score (P-F1)} and \emph{Mean Absolute Position Error (APS)} for tag insertion position, and \emph{Stability Character Error Rate (S-CER)} for transcription integrity after tag insertion. While the PRI method achieves a slight edge in tag positioning accuracy (P-F1 and APS), the PC-PTI approach demonstrates overwhelmingly superior stability, leaving the original transcription almost perfectly intact (S-CER of 0.35\% on our OV-Speech). Given that preserving the integrity of the ground-truth text is critical, we adopted the PC-PTI method for processing the entire OV-Speech dataset.

Through this five-stage pipeline, the OV-Speech dataset extends ContextSpeech with open-vocabulary instructions, reasoning chains, and fine-grained transcription with paralinguistic tags, providing a comprehensive and solid foundation for OV-InstructTTS.

\section{The OV-InstructTTS-TEP Framework}
\label{sec:method}

\begin{figure}[ht]
  \centering
  \includegraphics[width=\linewidth]{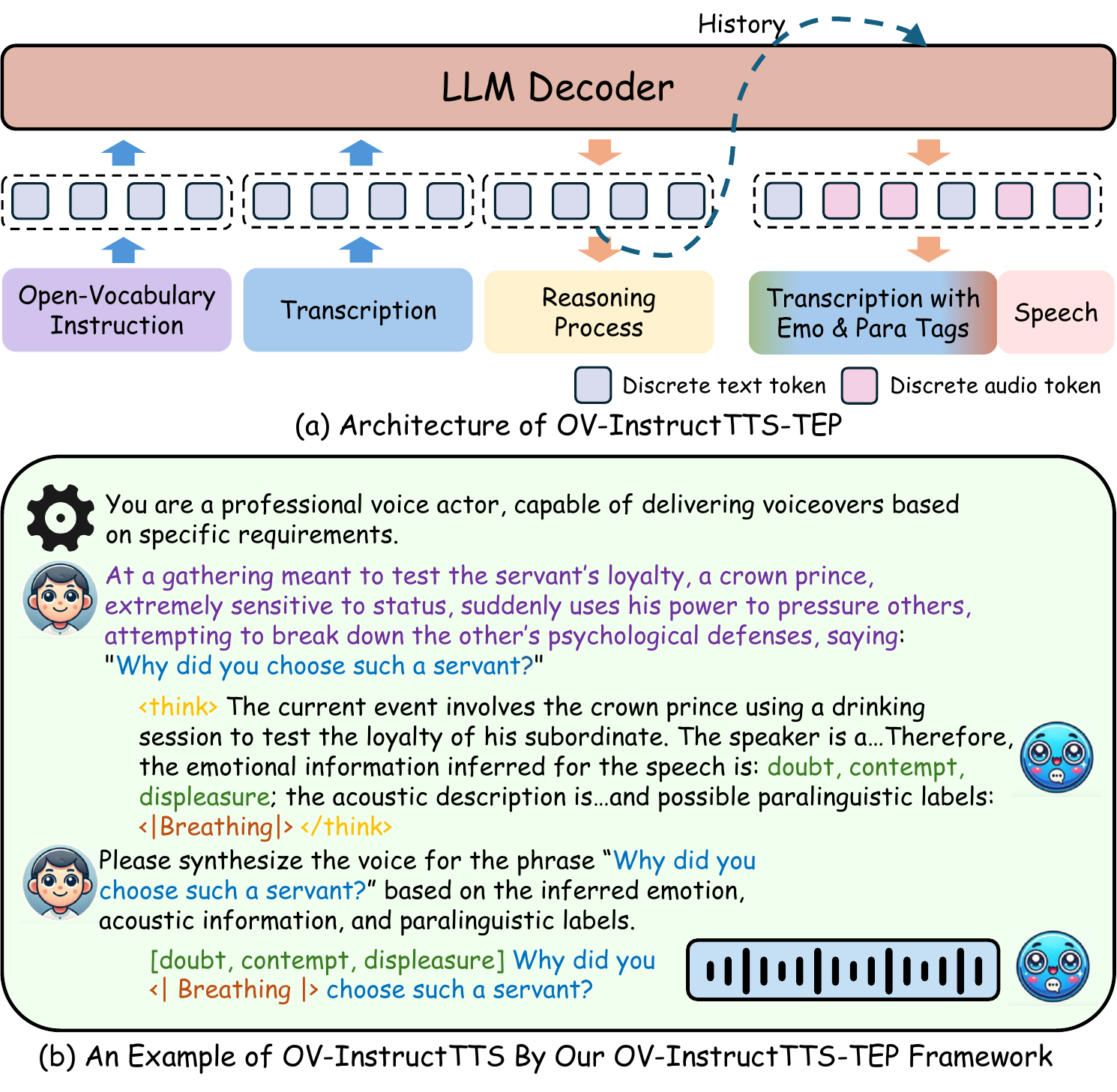}
  \vspace{-10pt}
  \caption{The architecture of our proposed reasoning-based OV-InstructTTS model and an example.}
  \label{fig:model}
\vspace{-15pt}
\end{figure}

We propose OV-InstructTTS-TEP, a reasoning-driven framework that leverages the unified comprehension, reasoning, and generation capabilities of LALMs for OV-InstructTTS.
This framework reformulates OV-InstructTTS from a direct high-level instruction-to-speech mapping into a structured two-step process. Rather than implicitly learning the mapping, the model is explicitly guided to first “think” by generating a reasoning chain, then synthesizing speech conditioned on it. This design leverages the strong in-context learning capabilities of LALMs to bridge the semantic gap between high-level user instructions and low-level acoustic realizations. We adopt Step-Audio-2-mini-Base\footnote{https://github.com/stepfun-ai/Step-Audio2} \cite{wu2025step} as our pretrained ALM backbone for its strong audio comprehension and open-source availability, and apply Supervised Fine-Tuning (SFT) on our OV-Speech dataset. As illustrated in Fig. \ref{fig:model}(a), our model takes an open-vocabulary instruction and a text transcript as input and proceeds in two steps:
\begin{itemize}[leftmargin=*]
\item \textbf{Thinking Token Generation:} The model first generates a textual reasoning process that maps the open-ended instruction to emotion labels, acoustic descriptions, and paralinguistic labels. It deconstructs the high-level instruction, identifying which of the contextual elements (environmental descriptions, the current event, the speaker’s personality, the interlocutor’s state, and the speaker’s intent) are present, and then infers the appropriate attributes for the speech performance in the given context.
\item \textbf{Interleaved Text-Audio Token Generation:} Conditioned on the preceding reasoning history, the model generates an interleaved sequence of text and audio tokens. The text tokens represent an enhanced version of the original transcript, augmented with the inferred attributes in the format: \texttt{[emotion label] Transcript with <|paralinguistic tags|>}. The discrete audio tokens are subsequently synthesized into a waveform using the same Flow-matching model and HiFiGAN vocoder as Step-Audio-2-mini-Base.
\end{itemize}

Fig. \ref{fig:model}(b) provides a concrete example. Given a complex situational instruction, our model first generates a \texttt{<think>} block that analyzes the context to infer emotion labels (\texttt{[doubt, contempt, displeasure]}), acoustic descriptions, and paralinguistic tags (\texttt{<|Breathing|>}). Subsequently, it predicts tokens of enhanced transcript (\texttt{[doubt, contempt, displeasure] Why did you <|Breathing|> choose such a servant?}) and the corresponding audio, ultimately synthesizing the speech. This reasoning-driven approach is the key to producing expressive speech that aligns with open-ended natural language descriptions, ensuring that the generated voice appropriately reflects nuanced emotional and paralinguistic cues beyond literal text content.

\begin{table*}[ht]
\caption{Overall performance comparison with baseline systems. \textcolor{deepred}{\textbf{Red}} highlights the best performance, and \textcolor{blue}{\textbf{blue}} indicates the second-best. $^\diamond$Indicates that the speaker's timbre was converted to the target speaker using the tokenizer and flow-matching of CosyVoice2.}
\label{tab:r1}
\centering
\resizebox{0.98\textwidth}{!}{
\begin{tabular}{lcccccc}
\toprule
 & Gemini Score\textcolor{red}{$\uparrow$} & Gemini Rank\textcolor{deepgreen}{$\downarrow$} & CER(\%)\textcolor{deepgreen}{$\downarrow$} & SIM\textcolor{red}{$\uparrow$} & MOS\textcolor{red}{$\uparrow$} & ICMOS\textcolor{red}{$\uparrow$} \\
\hline
\rowcolor{shadecolor}
GroundTruth                             & 75.43 & 2.94/6 & 3.10 & - & 4.10 ($_{\pm 0.14}$) & 4.33 ($_{\pm 0.15}$)\\
\cdashline{1-7}
Cosyvoice2 (No-Instruct)                & 66.99 & 3.59/6 & \textcolor{deepred}{\textbf{3.09}} & 0.659 & \textcolor{blue}{\textbf{3.84 ($_{\pm 0.19}$)}} & 2.94 ($_{\pm 0.23}$)\\
GPT4o$^\diamond$                        & \textcolor{blue}{\textbf{68.31}} & \textcolor{blue}{\textbf{3.48/6}} & 3.89 & 0.701 & 3.23 ($_{\pm 0.24}$) & 2.42 ($_{\pm 0.23}$)\\
Higgs Audio V2$^\diamond$               & 65.10 & 3.73/6 & 8.42 & \textcolor{blue}{\textbf{0.707}} & 3.81 ($_{\pm 0.20}$) & \textcolor{blue}{\textbf{3.00 ($_{\pm 0.20}$)}}\\
Step-Audio-2-mini                       & 67.59 & 3.56/6 & 5.49 & 0.701 & 3.53 ($_{\pm 0.24}$) & 2.40 ($_{\pm 0.21}$)\\
\rowcolor{LightCyan1}
\textbf{OV-InstructTTS-TEP}        & \textcolor{deepred}{\textbf{70.42}} & \textcolor{deepred}{\textbf{3.39/6}} & \textcolor{blue}{\textbf{3.61}} & \textcolor{deepred}{\textbf{0.722}} & \textcolor{deepred}{\textbf{4.28 ($_{\pm 0.14}$)}} & \textcolor{deepred}{\textbf{3.91 ($_{\pm 0.17}$)}}\\
\hline
\end{tabular}
}
\vspace{-12pt}
\end{table*}

\begin{table*}[h!]
\caption{Ablation study on the impact of key components: the (T)hinking process and the predicted token of enhanced transcriptions with (E)motional labels and (P)aralinguistic tags. \textcolor{deepred}{\textbf{Red}} highlights the best performance, and \textcolor{blue}{\textbf{blue}} indicates the second-best.}
\label{tab:r3}
\centering
\resizebox{0.98\textwidth}{!}{
\begin{tabular}{lcccccc}
\toprule
 & Gemini Score\textcolor{red}{$\uparrow$} & Gemini Rank\textcolor{deepgreen}{$\downarrow$} & CER(\%)\textcolor{deepgreen}{$\downarrow$} & SIM\textcolor{red}{$\uparrow$} & MOS\textcolor{red}{$\uparrow$} & ICMOS\textcolor{red}{$\uparrow$} \\
\hline
\rowcolor{shadecolor}
GroundTruth                     & 76.02 & 3.41/8 & 3.10 & - & 4.10 ($_{\pm 0.14}$) & 4.33 ($_{\pm 0.15}$)\\
\cdashline{1-7}
(a) Step-Audio-2-mini (No-Instruct) & 61.49 & 4.85/8 & 8.06 & 0.684 & 3.70 ($_{\pm 0.22}$) & 2.57 ($_{\pm 0.22}$)\\
(b) Step-Audio-2-mini               & 63.18 & 4.75/8 & 5.49 & 0.701 & 3.53 ($_{\pm 0.24}$) & 2.40 ($_{\pm 0.21}$)\\
\cdashline{1-7}
(c) TTS (No-Instruct)     & 66.34 & 4.48/8 & 3.78 & 0.715 & 4.15 ($_{\pm 0.21}$) & 3.61 ($_{\pm 0.20}$)\\
(d) OV-InstructTTS                   & 67.70 & 4.40/8 & \textcolor{blue}{\textbf{3.56}} & \textcolor{blue}{\textbf{0.720}} & 4.23 ($_{\pm 0.16}$) & 3.74 ($_{\pm 0.18}$)\\
(e) OV-InstructTTS-EP           & 66.98 & 4.45/8 & 3.65 & \textcolor{deepred}{\textbf{0.722}} & \textcolor{blue}{\textbf{4.27 ($_{\pm 0.16}$)}} & 3.81 ($_{\pm 0.16}$)\\
(f) OV-InstructTTS-T            & \textcolor{blue}{\textbf{68.71}} & \textcolor{blue}{\textbf{4.26/8}} & \textcolor{deepred}{\textbf{3.45}} & \textcolor{deepred}{\textbf{0.722}} & \textcolor{blue}{\textbf{4.27 ($_{\pm 0.18}$)}} & \textcolor{blue}{\textbf{3.90 ($_{\pm 0.16}$)}}\\
\rowcolor{LightCyan1}
(g) \textbf{OV-InstructTTS-TEP} & \textcolor{deepred}{\textbf{71.57}} & \textcolor{deepred}{\textbf{3.89/8}} & 3.61 & \textcolor{deepred}{\textbf{0.722}} & \textcolor{deepred}{\textbf{4.28 ($_{\pm 0.14}$)}} & \textcolor{deepred}{\textbf{3.91 ($_{\pm 0.17}$)}}\\
\hline
\end{tabular}
}
\vspace{-12pt}
\end{table*}

\section{Experiments}
\label{sec:exp}

\subsection{Experimental Setup}

\textbf{Datasets.}
We partition the proposed OV-Speech dataset into training and test sets. The training set consists of 316,807 utterances drawn from 83 novels, with each utterance paired with three distinct open-vocabulary instructions. The test set includes 1,500 utterances from 3 held-out novels, ensuring that both speakers and narrative contexts remain unseen during training.


\textbf{Implementation Details.}
We build our system upon the Step-Audio-2-mini-Base model. We finetune the model on OV-Speech using a learning rate of $1 \times 10^{-5}$ and a global batch size of 32. All experiments were conducted on 8 NVIDIA A100 (80GB) GPUs.

\textbf{Baselines.}
We compare our model against several high-performance TTS or LALM systems. GPT4o is evaluated via its official API. Higgs-Audio-V2\footnote{https://github.com/boson-ai/higgs-audio}, CosyVoice2\footnote{https://github.com/FunAudioLLM/CosyVoice}, and Step-Audio-2-mini\footnote{https://github.com/stepfun-ai/Step-Audio2} are evaluated using their publicly available codebases and checkpoints. We select these baselines for their strong performance: CosyVoice2 is a widely recognized TTS model known for synthesizing speech with high naturalness, while the other three models demonstrate notable instruction-following capabilities, enabling them to synthesize speech from open-vocabulary descriptions.

\subsection{Evaluation Metrics}

We evaluate performance from multiple perspectives:
\textbf{Instruction Following. }To evaluate how well the synthesized speech follows open-vocabulary instructions, we use an LLM-as-a-judge approach. Gemini \footnote{gemini-2.5-pro-preview-05-06} serves as the AI evaluator, where a set of speech samples is input, and Gemini assigns scores based on the alignment between the speech and the open-vocabulary instructions, providing both a \emph{Gemini Score} (0-100) and a \emph{Gemini Rank} within the group. 
\textbf{Intelligibility. }We measure \emph{Character Error Rate (CER)} using the ASR model Paraformerzh \cite{gao2023funasr} to evaluate speech intelligibility. 
\textbf{Timbre Similarity. } We compute \emph{cosine similarity (SIM)} between speaker embeddings from a WavLM-large \cite{chen2022wavlm} model to measure the timbre similarity. 
\textbf{Subjective Evaluation. } We further conduct human listening tests with 8 native Mandarin speakers. Listeners rate audio samples on two dimensions: the overall \emph{Mean Opinion Score (MOS)}, which reflects perceived naturalness and audio quality on a 5-point scale, and the \emph{Instruction Consistency MOS (ICMOS)}, which measures how faithfully the generated speech aligns with the given instruction, also on a 5-point scale.

\subsection{Main Results}
As shown in Table~\ref{tab:r1}, our proposed OV-InstructTTS-TEP achieves the best overall performance across both objective and subjective metrics. 
For the Gemini-based evaluation, our proposed model achieves the highest Gemini Score (70.42) and the best Gemini Rank (3.39/6), excluding GroundTruth. This demonstrates a substantial improvement over existing TTS approaches in open-vocabulary instruction following.
In subjective evaluations, it outperforms all baselines in both naturalness (MOS 4.28) and instruction consistency (ICMOS 3.91), with MOS even exceeding that of GroundTruth recordings. Furthermore, the model achieves the best speaker similarity (SIM 0.722) while maintaining a competitive CER (3.61\%), confirming its ability to generate clear and speaker-consistent speech. The strong performance across all these metrics validates the effectiveness of our method.

\subsection{Ablation Study}
To dissect the contribution of the OV-Speech dataset and each component in our OV-InstructTTS-TEP framework, we conducted a comprehensive ablation study, as summarized in Table~\ref{tab:r3}.

\textbf{OV-Speech dataset.} Comparing Step-Audio-2-mini without instruction fine-tuning (a,b) to models fine-tuned on our OV-Speech dataset (c,d), we observe clear gains: Gemini Score increases from 61.49 to 66.34 in the no-instruction setting and from 63.18 to 67.70 in the instruction setting. MOS and ICMOS also improve substantially. Moreover, within the fine-tuned models, the instruction-based variant (d) consistently outperforms its no-instruction counterpart (c), confirming the effectiveness of the OV-Speech dataset.

\textbf{Effect of reasoning.} Introducing an explicit reasoning step yields consistent improvements. Comparing (d) with (f), Gemini Score rises from 67.70 to 68.71, while MOS and ICMOS also improve (4.23→4.27 and 3.74→3.90, respectively). This supports our hypothesis that reasoning is crucial for bridging high-level instructions with low-level acoustic attributes, enabling the model to generalize beyond memorized associations.

\textbf{Synergy of reasoning and fine-grained transcriptions.} 
Interestingly, predicting transcription tokens with emotional labels and paralinguistic tags alone (e) does not always outperform the baseline (d), with the Gemini Score slightly decreasing (67.70 → 66.98). This may be due to the lack of reasoning, which affects the accuracy of emotion and tag predictions. However, when combined with reasoning (g), the model achieves the best results: the Gemini Score rises to 71.57, the Gemini Rank improves to 3.89/8, and both MOS and ICMOS also increase. This shows that reasoning enables the model to infer fine-grained attributes from open-vocabulary instructions, making the enriched transcriptions predicted more expressive and consistent with the instruction.

\section{Conclusion}
\label{sec:conclusion}
In this paper, we introduced OV-InstructTTS, a new paradigm that addresses the limitations of existing InstructTTS systems, where instructions are often confined to low-level acoustic attributes. To support this paradigm, we constructed OV-Speech, a dataset featuring open-vocabulary instructions, reasoning process annotation, and paralinguistic-aware transcriptions. Building on this resource, we further proposed OV-InstructTTS-TEP, a reasoning-driven framework that first “thinks” over the given instruction and then performs interleaved generation of enriched text and audio tokens before synthesizing speech. Extensive experiments demonstrate that our approach substantially outperforms baselines in instruction following, naturalness, and speaker similarity, with both reasoning and enriched text token prediction proving crucial. This work paves the way for more intuitive, flexible, and user-centric controllable TTS.

\vfill\pagebreak

\bibliographystyle{IEEEbib}
\bibliography{strings,refs}

\end{document}